             \documentstyle[12pt,epsf]{article}
\begin{document}\begin{flushright}\thispagestyle{empty}
ESI--619\\
OUT--4102--77\\
MZ--TH/98--42\\
hep-th/9810087\\
13 October 1998                \end{flushright}\vspace*{2mm}\begin{center}{
                                                    \Large\bf
Renormalization automated by Hopf algebra           }\vglue 10mm{\large{\bf
D.~J.~Broadhurst$^{1)}$ and D.~Kreimer$^{2)}$       }\vglue 4mm
Erwin Schr\"odinger Institute, A-1090 Wien, Austria }\end{center}\vfill
                                                  \noindent{\bf Abstract}\quad
It was recently shown that the renormalization of quantum field theory is
organized by the Hopf algebra of decorated rooted trees, whose coproduct
identifies the divergences requiring subtraction and whose antipode achieves
this. We automate this process in a few lines of recursive symbolic code, 
which deliver a finite renormalized expression for any Feynman diagram. We 
thus verify a representation of the operator product expansion, which 
generalizes Chen's lemma for iterated integrals. The subset of diagrams whose 
forest structure entails a unique primitive subdivergence provides a 
representation of the Hopf algebra ${\cal H}_R$ of undecorated rooted trees. 
Our undecorated Hopf algebra program is designed to process the 24,213,878 
BPHZ contributions to the renormalization of 7,813 diagrams, with up to 12 
loops. We consider 10 models, each in 9 renormalization schemes. The two 
simplest models reveal a notable feature of the subalgebra of Connes and 
Moscovici, corresponding to the commutative part of the Hopf algebra 
${\cal H}_T$ of the diffeomorphism group: it assigns to Feynman diagrams 
those weights which remove zeta values from the counterterms of the minimal 
subtraction scheme. We devise a fast algorithm for these weights, whose 
squares are summed with a permutation factor, to give rational counterterms.
\vfill\footnoterule\noindent
$^1$) D.Broadhurst@open.ac.uk;
http://physics.open.ac.uk/$\;\widetilde{}$dbroadhu\\
permanent address: Physics Dept, Open University, Milton Keynes MK7 6AA, UK\\
$^2$) Dirk.Kreimer@uni-mainz.de;
http://dipmza.physik.uni-mainz.de/$\;\widetilde{}$kreimer\\
Heisenberg Fellow, Physics Dept, Univ.\ Mainz, 55099 Mainz, Germany
\newcommand{\bookfig}[5]{\begin{figure}\centering\fbox{%
\epsfysize=#5cm\epsfbox{#1}}\caption[#2]{\small #4}\label{#3}\end{figure}}
\newcommand{\ep}{\varepsilon}\newcommand{\de}{\delta}
\newpage\setcounter{page}{1}

\section{Introduction}

Perturbative quantum field theory~(pQFT) entails the process
of renormalization: the iterated subtraction of subdivergences of
Feynman diagrams, culminating in the subtraction of an overall
divergence (or addition of a counterterm) that renders each diagram finite
and delivers its contribution to the appropriate renormalized Green function.
Many people helped to develop this process, though it has
become customary to record the work of Bogoliubov and Parasiuk~\cite{BP},
Hepp~\cite{H}, and Zimmermann~\cite{Z} (BPHZ), completed by the forest
formula in~\cite{Z}, almost 30 years ago.
A review is provided by the textbook of Collins~\cite{C}.
In addition to the BPHZ formalism one needs analysis,
to evaluate regularized expressions for the bare diagrams, prior
to renormalization. Much progress has been made in this direction
by combining dimensional regularization~\cite{DR} and integration by
parts, for massless~\cite{CT} and massive~\cite{DJB,LVA,sixth}
diagrams with up to 3 loops, together with more {\em ad hoc\/} techniques
at higher loops. Since 1981 it was possible, in principle,
to obtain the 4-loop $\beta$-function of QCD by combining the BPHZ
formalism with the algorithm of~\cite{CT} for massless 3-loop
2-point functions in $D:=4-2\ep$ spacetime dimensions. Yet not until
very recently was a result obtained in~\cite{beta}, and even there
the BPHZ formalism was not fully exploited. This serves as an indication
of the challenge of {\em organizing\/} renormalization.

In this paper, we exploit the work in~\cite{DK,CK,DKover,OPE},
which shows that the joblist of renormalization is
encapsulated by the coproduct,
$\Delta$, of the Hopf algebra of decorated rooted trees
and that counterterms are given by its
antipode, $S$. We show how truly simple it is to automate renormalization
by recursive definitions of $\Delta$ and $S$ and hence to
calculate the contribution of any diagram
to a renormalized Green function, in any situation where analytic
methods are adequate to evaluate the regularized bare diagrams
entailed by $\Delta$. We emphasize that in this work we use only well-tried
analytic methods. Our aims are to make transparent the Hopf algebra
of renormalization, to program it efficiently, to apply the
program to cases where analysis is possible at large numbers of loops,
and to report the first fruit of
the process of discovery that is thereby enabled.

In sect.~2 we review 4 crucial formul{\ae} from~\cite{DK,CK}, which
capture the entirety of renormalization, and rewrite them as a few lines of
code in the symbolic manipulation language~{\sc Reduce}~\cite{red},
whose translation to other languages should present little difficulty
to readers with other preferences. In sect.~3 we give a representation
of the operator product expansion, recently found by DK~\cite{OPE}
in the course of extending Chen's lemma~\cite{Chen} on iterated integrals,
and hence providing a powerful test of the correctness of our code.
In sect.~4 we specialize to the Hopf algebra of undecorated rooted trees,
where analytic methods and efficient programming allow us to study
its 7,813 Feynman diagrams with loop numbers $n\le12$ in a wide
variety of field theories and renormalization schemes. In sect.~5
we report our first discovery with this new tool, by showing the
remarkable simplification that results when one combines diagrams
with the weights specified by Connes and Moscovici~\cite{CM}, in their
study of the subalgebra entailed by the diffeomorphism group.
Sect.~6 offers conclusions and suggestions for further study.

\section{Renormalization by Hopf algebra}

The divergence structure of a Feynman diagram is naturally
represented by a tree~\cite{CK}, whose vertices (or nodes)
represent primitive
divergences, with edges connecting vertices in a manner that
encodes the nesting (or forest structure) of subdivergences.
Each primitive divergence is associated with a skeleton diagram~\cite{BD},
free of subdivergences. To distinguish these we need labels, $\gamma_k$.
Fig.~1 gives an example from QED, entailing one-loop skeleton diagrams,
$\gamma_1$ and $\gamma_3$, in the electron and photon propagators,
and one- and two-loop skeleton diagrams, $\gamma_0$ and $\gamma_2$,
in the coupling. The assignment of labels is arbitrary, so long
as it permits no confusion. Now consider the 6-loop diagram
of Fig.~1. How shall we encode its divergence structure?

\bookfig{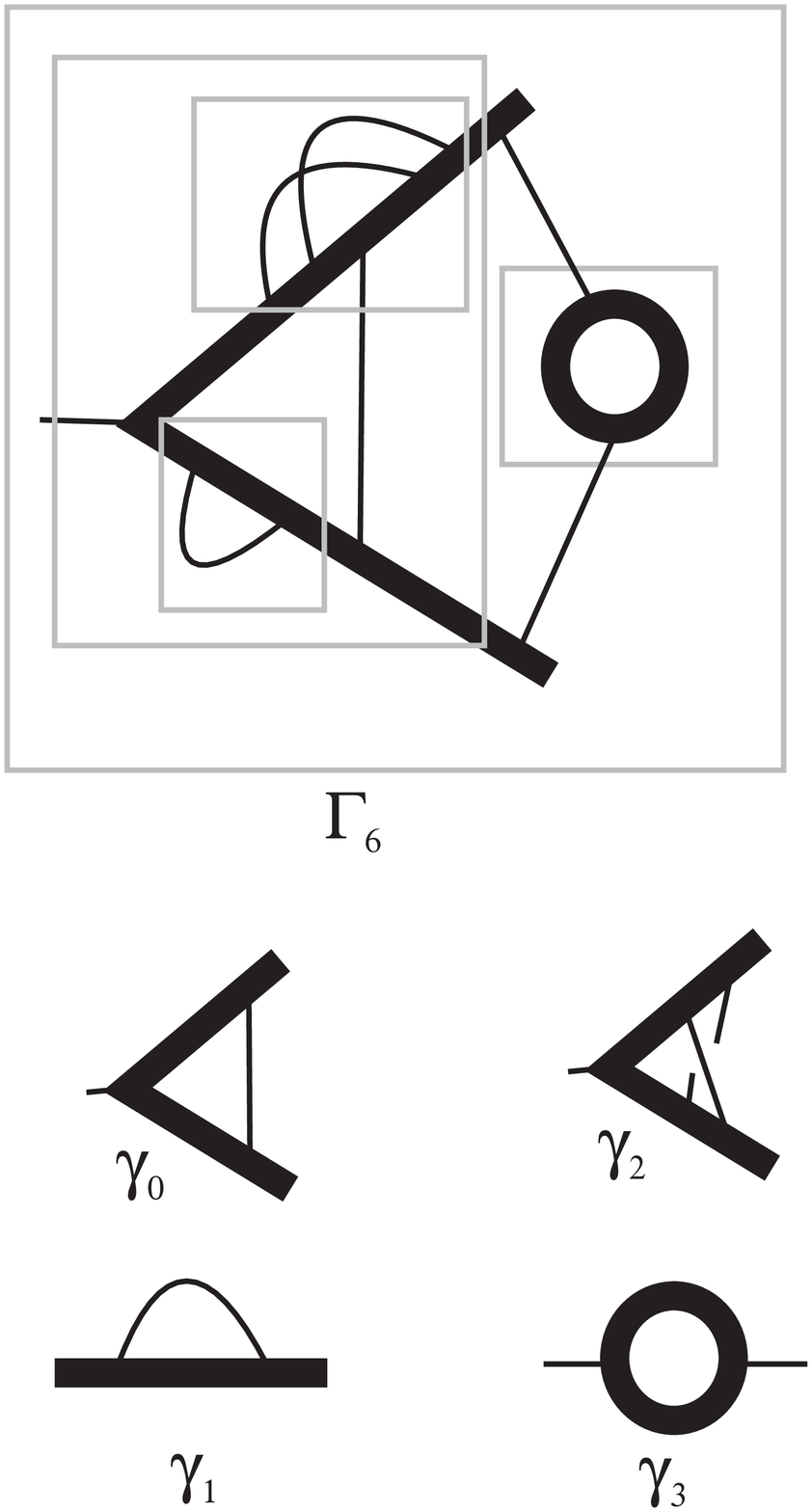}{6loops}{ddf1}{The divergences in $\Gamma_6$
are generated by 4 skeleton diagrams.}{8}

One obvious method is to form a list\footnote{In~\cite{DK}
the equivalent information was encoded by parentheses, written
in the reverse order. For present purposes, it is more efficient
to work with lists, with the outermost label at the head.}:
\begin{equation}
\Gamma_6=\{0,\{3\},\{0,\{1\},\{2\}\}\}\label{F1}
\end{equation}
The first element of the list is a label, which tells us in this
case that the subdivergences reside inside the one-loop skeleton
for the coupling.
The rest of the list consists of sublists. The first of these
is $\{3\}$, which encodes the one-loop primitive divergence in the
photon propagator; the second is $\{0,\{1\},\{2\}\}$,
which we parse in exactly the same manner as before: it tells us
that inside a one-loop coupling skeleton there reside a one-loop
electron-propagator skeleton and a two-loop coupling skeleton.
It is also rather convenient that this iterated structure,
list=head+sublists, allows us to preserve the order of sublists,
which may later be useful in taking traces over spins.

An equally obvious, and entirely equivalent, coding is provided
by the rooted tree of Fig.~2. It grows (downwards!) from
the root 0, which has a pair of branches. Each of these branches
is itself a rooted tree: the first has root 3, but no branches; the second
has root 0 and branches which are the branchless trees with roots 1 and 2.
Clearly, there is a trivial translation between the structure
tree=root+branches and the structure list=head+sublists in~(\ref{F1}).
In each case there is a feature which encapsulates the forest structure of
divergences: each branch is itself a tree; each sublist is itself
a list.

\bookfig{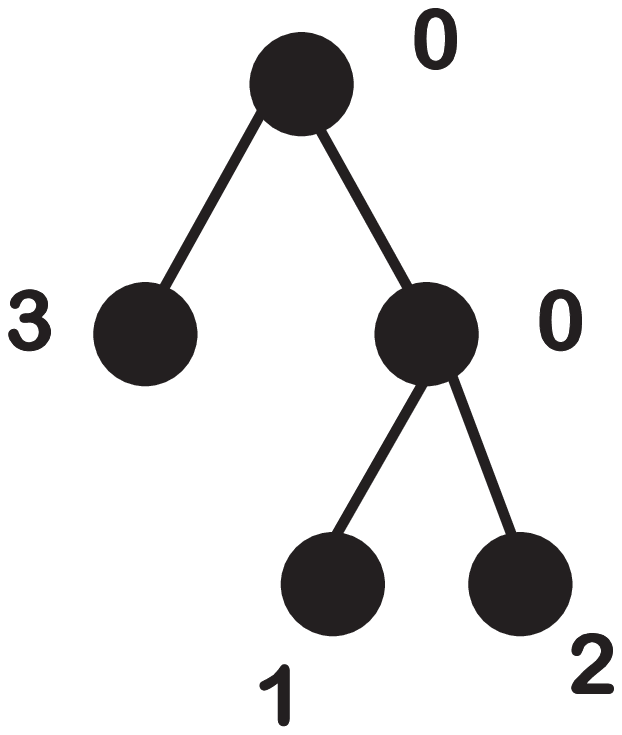}{Tree}{ddf2}{Divergences encoded by a rooted
tree, corresponding to the list~(1).}{4}

We shall present the structure of the Hopf algebra in terms
of trees, roots, and branches (which are trees); we shall encode it
in terms of lists, heads, and sublists (which are lists).
In~\cite{CK,DKover}
it is shown that overlapping divergences are accommodated
by this iterative structure, since they require, at most, a sum
of lists of decorations.

\subsection{Coproduct}

Let $a\in{\cal A}$ be a tree, where ${\cal A}$ is the algebra
of trees, with a product that corresponds merely to the commutative
product of the associated Feynman diagrams. A coproduct $\Delta(a)
\in{\cal A}\otimes{\cal A}$, which is not cocommutative,
is defined recursively by~\cite{DK,CK}
\begin{equation}
\Delta(a)=a\otimes e+id\otimes B^+_r(\Delta(B^-(a)))\label{D}
\end{equation}
where $e$ is the empty tree (evaluating to unity); $id$ is the
identity map in ${\cal A}$; $r$ is the root
of $a$; $B^-(a)\in{\cal A}$ is the product of the branches of $a$;
and $B^+_r$ is the operator that maps any
element $c\in{\cal A}$ to the tree with root $r$ and branches $c$.
The recursion terminates with
\begin{equation}
\Delta(e)=e\otimes e\label{De}
\end{equation}
and is effected by applying the product rule
\begin{equation}
\Delta(b)=\mbox{$\prod_k$}\Delta(b_k)
\end{equation}
to the product of branches $b=\prod_k b_k$ obtained by
removing the root $r$ of $a$, and then recombining the products on the
right of the tensor product with the original root $r$.

It was shown in~\cite{DK} that $\Delta(a)$ generates
the joblist of the practical field theorist engaged upon
the renormalization of a Feynman diagram with divergence structure given
by the tree $a$. On the left of the tensor product reside (products of)
subdiagrams requiring counterterms; on the right reside diagrams in which
these subdiagrams shrink to points.
Most of us will testify, if pressed, that
it is perilously easy to forget where one has reached when subtracting
subdivergences by hand.
On the other hand, a correctly programmed recursive procedure,
in a suitable symbolic manipulation language,
will not lose track. An implementation
in {\sc Reduce} is provided by
\begin{verbatim}
   for all a,b such that first a=af let A2 a*A2 b=A2 append(a,{b});
   procedure D a;
   A1 a+sub(af=first a,A2{af}*for k:=2:length a product D part(a,k));
\end{verbatim}
where {\tt A1} and {\tt A2} are operators that hold lists
on the left and right of ${\cal A}\otimes{\cal A}$ and {\tt af}
is a dummy argument that is replaced by {\tt first a} after
each recursion. The {\tt let} statement in the first line sweeps
up products on the right of ${\cal A}\otimes{\cal A}$.
In the body of the procedure,
the list representing the tree is beheaded by starting at {\tt k:=2};
the product of coproducts of sublists is taken; then
product terms on the right are reheaded by the original root.
This generates the
joblist of renormalization, which finally reads,
in Sweedler notation~\cite{CK}
\begin{equation}
\Delta(a)=\mbox{$\sum_k$} a_k^{(1)}\otimes a_k^{(2)}\label{sweedler}
\end{equation}
with products of trees on the left, and single trees on the right,
of the tensor product.

\subsection{Antipode}

An antipode, $S$, upgrades the bi-algebra
defined by $\Delta$ to a Hopf algebra. For any tree $a\in{\cal A}$,
the antipode, $S(a)\in{\cal A}$, is given, again recursively,
by~\cite{DK,CK}
\begin{equation}
S(a)=-a-\mbox{$\sum_k^\prime$}S(a_k^{(1)})a_k^{(2)}\label{S}
\end{equation}
where $\sum^\prime_k$ omits the terms $a\otimes e$
and $e\otimes a$ of the full Sweedler sum~(\ref{sweedler}). The encoding is
\begin{verbatim}
   procedure S a; sub(A1=S,A2=A1,A1 a-D a);
\end{verbatim}

\subsection{Counterterm}

Renormalization consists in applying an operator $R$, on the left,
at each recursion in the computation of the antipode.
In the momentum (MOM) scheme,
$R_{\rm MOM}$
is merely the instruction to replace the external momenta
by a set of fiducial momenta. In the
minimal subtraction (MS) scheme, $R_{\rm MS}$ nullifies external
momenta and internal masses and takes only the singular terms
of the Laurent expansion in $\ep$.
In either case, we have
\begin{eqnarray}
R(R(x))&=&R(x)\label{idemp}\\
R(R(x)R(y))&=&R(x)R(y)\label{rprod}\\
R(x+y)&=&R(x)+R(y)\label{rlin}
\end{eqnarray}
The counterterm, associated by scheme $R$ to
a Feynman diagram with divergence structure given by tree $a$,
is~\cite{DK,CK}
\begin{equation}
S_R(a)=-R(a)-\mbox{$\sum_k^\prime$}R(S_R(a_k^{(1)})a_k^{(2)})\label{S_R}
\end{equation}
and is hence encoded by
\begin{verbatim}
   procedure S_R a; R(sub(A1=S_R,A2=A1,A1 a-D a));
\end{verbatim}

\subsection{Renormalized Green function}

The Green-function contribution
of a diagram with divergence structure $a$ is
\begin{equation}
\Gamma_R(a)=\lim_{\ep\to0}\mbox{$\sum_k$}S_R(a_k^{(1)})a_k^{(2)}\label{G_R}
\end{equation}
and is hence encoded by
\begin{verbatim}
   procedure G_R a; sub(A1=S_R,A2=A1,D a);
\end{verbatim}
which generates $2^n$ terms in the case of a tree with $n$ vertices, each
representing a primitive divergence. The limit $\ep\to0$ is not specified
in the procedure, but is guaranteed to exist when one provides concrete
expressions for the dimensionally regularized bare diagrams, as we
shall do in sect.~4.

\subsection{Program {\tt handsgam.red}}

The {\sc Reduce} program {\tt handsgam.red}, available\footnote{See {\tt
ftp://physics.open.ac.uk/pub/physics/dbroadhu/ha/hareadme.txt}
for details.} via FTP, adds 5 features.
\begin{enumerate}
\item A {\tt noncom} instruction prevents reordering of sublists. This is
useful in cases where one needs to take account of spin structure.
\item The linearity of $R$ is declared by {\tt let}
statements.
\item The loop number of a tree is computed by adding the loop numbers
of its primitive decorations. This
is included as an argument of $R$, to facilitate Laurent
expansion in the MS scheme.
\item Results for $\Delta$, $S_R$ and $\Gamma_R$ are stored, to speed
up batch processing of diagrams with
subdivergences in common, and tests of the operator product expansion.
\item A test is made for the arbitrarily decorated tree
$\{1,\{2\},\{3,\{4\},\{5\}\},\{6\}\}$,
to ensure that the program correctly generates the 64 terms of the
BPHZ formalism.
\end{enumerate}
The reader is invited to generate the 64 BPHZ terms by hand,
and then to estimate how long it might take to
generate 24,213,878 BPHZ terms for the 7,813 diagrams of sect.~4, without
the assistance of Hopf algebra.

\section{Chen's lemma and the OPE}

Suppose we renormalize in the MOM scheme at some
fiducial momenta labelled generically by $p_1$ and evaluate
$\Gamma_{\rm MOM}$ at some different external momenta $p_2$. Let
$\Gamma_{\rm MOM}^{(1,2)}(a)$
denote the result for a Feynman diagram with divergence structure $a$.
In~\cite{OPE} it is shown that
\begin{equation}
\Gamma_{\rm MOM}^{(1,2)}(a)=
\mbox{$\sum_k$}\Gamma_{\rm MOM}^{(1,0)}(a_k^{(1)})
\Gamma_{\rm MOM}^{(0,2)}(a_k^{(2)})\label{ope}
\end{equation}
where the sum is over the Sweedler decomposition of the coproduct of $a$,
the momenta $p_0$ are arbitrary, and it is understood that
\begin{equation}
\Gamma_R(b)=
\mbox{$\prod_j$}\Gamma_R(b_j)
\end{equation}
for a product of trees $b=\prod_j b_j$, in any scheme $R$.

Identity~(\ref{ope}) is readily
obtained in the toy model of~\cite{DK}, which represents
the Hopf algebra of undecorated trees by iterations of products of integrals.
In that particular case, it corresponds to an extension of Chen's
lemma~\cite{Chen} for iterated integrals. It further extends
to arbitrarily decorated trees, in any field theory.
In~\cite{OPE} it is construed as a representation of Wilson's
operator product expansion~\cite{KW} (OPE),
whose distinctive feature is to allow one to express matrix elements
as sums of products with an arbitrary intermediate scale.
Here we use~(\ref{ope}) as a stringent test of our code:
program {\tt hamomope.red} verifies this MOM-scheme OPE.

There is a further result, for the MS scheme.
Consider a Feynman diagram, with divergence structure $a$, contributing
to a massless two-point function, with a single external
momentum\footnote{We adopt a metric such that $p^2>0$ in the
spacelike region.} $p$. Then
the MS-renormalized result will be a polynomial,
$\Gamma_{\rm MS}(a,L_\mu)$, in
$L_\mu:=\log(\mu^2/p^2)$, with degree equal to the loop number, where
$\mu$ is the renormalization scale of the MS scheme.
The
MOM-renormalized result will be a different polynomial,
$\Gamma_{\rm MOM}(a,L_1)$, in
$L_1:=\log(p_1^2/p^2)$, where $p_1$ is the fiducial momentum. However,
there is a strong connection between the sets of MS and MOM
polynomials for the terms in the Sweedler decomposition of the
coproduct $\Delta(a)$, namely
\begin{equation}
\Gamma_{\rm MS}(a,L_\mu-L_1)=
\mbox{$\sum_k$}
\Gamma_{\rm MS}(a_k^{(1)},L_\mu)
\Gamma_{\rm MOM}(a_k^{(2)},-L_1)\label{MSope}
\end{equation}
with the MOM-scheme results serving to transform MS-scheme results,
at momentum $p$, to a MS-scheme result with momentum $p_1$.

We shall apply~(\ref{MSope}) as a check of our methods in sect.~4.
After checking that the calculus passes this test, one
may economize by computing the MS results at $p^2=\mu^2$,
since their dependence on external momentum, determined by the
renormalization group, is generated by the MOM-scheme results,
by setting $L_\mu=0$ in~(\ref{MSope}).

\section{Hopf algebra of undecorated rooted trees}

We now specialize to the Hopf algebra of undecorated rooted trees,
generated by a single primitive divergence. From the point of view of
pQFT, this is a drastic step: the renormalized Green functions of the
theory, at a given order $n$,
are obtained by computing all decorations with total loop number $n$
and then combining them with the weights given by Wick contractions
of the perturbative expansion;
little interest might seem to attach to a set of Feynman diagrams
with only one-loop primitive propagator subdivergences.

In fact there are (at least) 6 cogent reasons for studying such diagrams.
\begin{enumerate}
\item Analytic results for such bare diagrams are immediately available,
in terms of highly structured products and quotients of
\begin{equation}
\Gamma(1-z)=\exp\left(\gamma z+\mbox{$\sum_{s>1}$} \zeta(s)z^s/s\right)
\label{Euler}
\end{equation}
where $z$ is an integer multiple of $\ep$, $\gamma$ is Euler's
constant, and $\zeta(s):=\sum_{k>0}1/k^s$ is the Riemann zeta function.
Hence we shall be able to obtain explicit results, of an analytically
nontrivial character, for 7,813 diagrams with loop numbers $n\le12$,
entailing the irreducible zeta values $\{\zeta(3),\zeta(5),\zeta(7),\zeta(9),
\zeta(11)\}$ and powers of $\pi^2$ from, for example, $\zeta(12)=691
\pi^{12}/638512875$. At the very least, this provides a healthy workout for
our Hopf algebra code. More importantly it enables us to look for structure
in the counterterms, lacking in the singularities of bare diagrams.
\item The 12 pure rainbow diagrams in our set of 7,813 correspond
to trees in which no vertex has fertility greater than unity. They
have a remarkable feature, observed in~\cite{DKknot} and proved to all orders
in~\cite{RD1,RD2}: their MS counterterms are rational polynomials in
$1/\ep$, bearing no trace of the derivatives of $\log\Gamma(1-z)$ at
$z=0$, despite the appearance of odd zeta values and powers of $\pi^2$
at every step of the BPHZ formalism. This potent example
of the simplicity of counterterms was the stimulus for the
investigations of~\cite{BKP,BDK,DKHab,BGK,BK15,knotsub,BKot,DKRhein},
where we discovered remarkable features of the nexus of knot/number/field
theory~\cite{DKbook}.
\item Trees in which only the root has fertility greater than zero
come from pure chains of one-loop self-energy divergences. In the case of
QED these generate the ultra-violet
Landau singularity of the photon propagator.
In the case of heavy-quark effective theory (HQET) they produce a
renormalon structure~\cite{BG} that frustrates unambiguous
Borel resummation of the perturbation series.
\item We may probe even deeper into the number-theory content of
the complex of all nestings of rainbows
and chains, constituting the full Hopf algebra of undecorated rooted trees,
by embedding it inside a two-loop diagram, as was done merely for
chains in~\cite{DJBchain}. Then we will encounter irreducible
multiple zeta values~\cite{BBB,DJBEul,BBBLa,BBBLc}
of the types $\zeta(5,3):=\sum_{m>n>0}1/m^5n^3$ and
$\zeta(3,5,3):=\sum_{k>m>n>0}1/k^3m^5n^3$, associated by pQFT to
the unique positive 3-braid 8-crossing knot and the unique
positive 4-braid 11-crossing knot, in the 7-loop analysis of~\cite{BKP}.
\item Next, and most excitingly, this seemingly trivial set of Feynman
diagrams was shown in~\cite{CK} to provide a solution of a universal
problem in Hochschild cohomology.
\item Finally, and most intriguingly, there is a unique combination
of these diagrams at loop number $n$ which represents a grading
of the commutative
part of the Hopf algebra ${\cal H}_T$ of the diffeomorphism group~\cite{CK}.
Specifically, if one adds the diagrams with integer weights
derivable from work by Connes and Moscovici~\cite{CM},
one expects to see some simpler structure,
the details of which neither they nor we anticipated. Thanks to the
discovery machine developed in this section, we shall have an interesting
result to report in sect.~5, with implications for the relation
between noncommutative geometry and quantum field theory.
\end{enumerate}

\subsection{Enumeration of rooted trees}

The numbers, $N(n)$, of rooted trees with $n$ vertices form a
sequence beginning~\cite{EIS}
\begin{verbatim}
   1,1,2,4,9,20,48,115,286,719,1842,4766,12486,32973,87811,235381,634847
\end{verbatim}
which grows exponentially, giving
\begin{verbatim}
   N(250) = 517763755754613310897899496398412372256908589980657316271
            041790137801884375338813698141647334732891545098109934676
\end{verbatim}
trees with 250 vertices. Already one sees a nontrivial feature of the
iterated structure: tree=root+branches, with every branch being itself a tree.
The asymptotic growth
\begin{equation}
N(n)={b\over n^{3/2}}c^n(1+O(1/n))\label{asy}
\end{equation}
entails constants
\begin{verbatim}
   b = 0.43992401257102530404090339143454476479808540794011
         98576534935450226354004204764605379862197779782334...
   c = 2.95576528565199497471481752412319458837549230466359
         65953504724789059647331395749510866682836765813525...
\end{verbatim}
that do not appear to be algebraic numbers, since
neither the PSLQ~\cite{PSLQ} nor the LLL~\cite{LLL} algorithm
was able to find rational polynomials that fit the first 100 digits
and correctly predict the next 10. The origin of their
(presumed) transcendentality resides in the functional equation
\begin{equation}
\log A(x)-A(x)=\log x+\sum_{k>1}{A(x^k)\over k}\label{fr}
\end{equation}
for the generating function $A(x):=\sum_{n>0}N(n)x^n$, which develops
a square-root branchpoint at $x=1/c$, near to which
$1-A(x)\sim\sqrt{1-c x}$. Setting $x=1/c$ we obtain
\begin{equation}
\log c=1+\sum_{k>1}{A(c^{-k})\over k}\label{cis}
\end{equation}
which allows one to determine $c$, by iterative solution, to an accuracy
of about $1/N(250)=O(10^{-114})$, after determining
$\{N(n)\mid n\le250\}$ from~(\ref{fr}), at small $x$. Then Taylor
expansion at $x=1/c$ yields
\begin{equation}
2\pi b^2=1+\sum_{k>1}{A^\prime(c^{-k})\over c^k}\label{bis}
\end{equation}
with the regularity of the r.h.s.\ of~(\ref{fr}) forcing nonlinear
relations between the coefficients, $c_k$, of
$A(x)=1-\sum_{k>0}c_k(1-c x)^{k/2}$.

We hope that these asymptotic results may help
colleagues to identify the Lie algebra dual to the Hopf algebra
${\cal H}_R$ of rooted trees, which solves a universal problem
in Hochschild cohomology~\cite{CK}.

\subsection{A variety of models}

We shall consider 10 models that generate values for the
bare diagrams: 6 of these correspond to well-defined field theories;
the remaining 4 are instructive toy models that probe the analytic
structure of bare and renormalized diagrams, while ignoring
fine details produced by spin and gauge dependence.

\subsubsection{BPHZ model}

In~\cite{DK} a BPHZ toy model was developed, by
iteration of products of integrals. To each tree, $a$, it assigns
a dimensionally regularized bare value, $B(a)$,
obtained by the recursion
\begin{equation}
B(a)=B_n(\ep)\mbox{$\prod_k$}B(b_k)\label{brec}
\end{equation}
where $n$ is the number of vertices of $a$; $b_k$ are the trees
formed from its branches; and
\begin{equation}
B_n(\ep)=\frac{L(\ep,n\ep)}{n\ep}\label{model}
\end{equation}
where $L(\ep,\de)$ is regular at both $\ep=0$ and $\de=0$.
This structure is generic; it covers all the field theories that
we study. The BPHZ model is a toy only because of the simplicity
of its master function, which is
\begin{equation}
L_{\rm BPHZ}(\ep,\de)=\frac{\pi\de}{\sin\pi\de}\lambda^{-\ep}
\label{BPHZ}
\end{equation}
resulting from nullifying all dimensionful parameters
except for an infra-red regulator, $\lambda$, in the outermost integration.

\subsubsection{Heavy-quark model}

More realistically, consider the propagator of a heavy quark in HQET,
at virtuality $\omega<0$. Neglecting a rational function of $\ep$ and
$\de$, which we later restore, the master function is
\begin{equation}
L_{\rm HQ}(\ep,\de)={\Gamma(1-\ep)\Gamma(1+2\de)\over
\Gamma(1-2\ep+2\de)}(-2\omega)^{-2\ep}\label{HQ}
\end{equation}
whose $\Gamma$ functions are immediately apparent in~\cite{EFT}.

\subsubsection{Covariant QFT model}

For a massless-quark propagator in QCD, with spacelike momentum $p$,
the $\Gamma$ functions in~\cite{CT} give
\begin{equation}
L_{\rm QFT}(\ep,\de)={\Gamma(1-\ep)
\Gamma(1-\de)\Gamma(1+\de)\over
\Gamma(1-\ep-\de)\Gamma(1-\ep+\de)}(p^2)^{-\ep}\label{QFT}
\end{equation}
with a rational factor which will be supplied later.

\subsubsection{MZV model}

To get multiple zeta values~\cite{BK15} (MZVs) in the bare diagrams,
we may embed the full complex of chains and rainbows
in a two-loop diagram, as was done for pure QED chains in~\cite{DJBchain}.
Here, we embed it in a finite diagram, so as to
remain within the undecorated Hopf algebra. This corresponds to using
the recursive process~(\ref{brec},\ref{model},\ref{QFT}), to compute an
$n$-loop bare value for a covariant propagator diagram, and then
multiplying by
\begin{equation}
L_n(\ep):=\frac{2L_{\rm QFT}(\ep,(n+1)\ep)
L_{\rm QFT}(\ep,(n+2)\ep)}{(n+1)(n+2)\ep^2}
S(0,-\ep,(n+1)\ep,-(n+2)\ep)
\label{MZV}
\end{equation}
where eq.~(17) of our paper with John Gracey~\cite{BGK} gives the
general result for the 2-loop 2-point function
$S(a,b,c,d)$, which generates MZVs via a ${}_3F_2$ hypergeometric series.
With $n_1+n_2=n_3+n_4$, we have
$S(n_1\ep,n_2\ep,n_3\ep,n_4\ep)=3(n_1n_2-n_3n_4)\ep^2\zeta(3)
+O(\ep^3)$, and hence obtain irreducible MZVs at level $n+2$
in the MS counterterms from undecorated trees with $n\ge6$ loops.
At $n=9$ we probe as deeply into the relation between knots and numbers
as we did in~\cite{BKP}; at $n=10$, we reach the 12-crossing knots
of~\cite{BGK,BK15}, which revealed an unexpected connection between
MZVs and alternating Euler sums~\cite{DJBEul}.

\subsubsection{Field theories}

As already indicated, we have only to supply a rational function
of $\ep$ and $\de$ to convert~(\ref{HQ}) to the realistic case
of a heavy-quark propagator in HQET.
In an arbitrary covariant gauge, with a gluon propagator
$g_{\mu\nu}/q^2+(\chi-1)q_\mu q_\nu/q^4$, we obtain
\begin{equation}
L_{\rm HQET}(\ep,\de)=\frac{3-2\ep-(1-2\ep)\chi}{1-2\de}
L_{\rm HQ}(\ep,\de)\label{HQET}
\end{equation}
which is divergence-free in Yennie gauge, with $\chi=3$.
We may compute in an arbitrary gauge, and also specialize to
\begin{equation}
L_{\rm HQF}(\ep,\de)=\frac{2}{1-2\de}
L_{\rm HQ}(\ep,\de)\label{HQF}
\end{equation}
in Feynman gauge, with $\chi=1$, and
\begin{equation}
L_{\rm HQL}(\ep,\de)=\frac{3-2\ep}{1-2\de}
L_{\rm HQ}(\ep,\de)\label{HQL}
\end{equation}
in Landau gauge, with $\chi=0$.

Similarly, we can convert~(\ref{QFT}) to the realistic case of
a light-quark propagator in QCD. The master function vanishes, identically,
in Landau gauge. Hence we compute in Feynman gauge, with
\begin{equation}
L_{\rm QCD}(\ep,\de)=-\frac{2(1-\ep)(1-\de)}
{(1-\ep-\de)(2-\ep-\de)}
L_{\rm QFT}(\ep,\de)\label{QCD}
\end{equation}
For the fermion propagator in Yukawa theory we have
\begin{equation}
L_{\rm Yuk}(\ep,\de)=-\frac{2(1-\de)}
{(1-\ep-\de)(2-\ep-\de)}
L_{\rm QFT}(\ep,\de)\label{YUK}
\end{equation}
and for $\phi^3$ theory, in $6-2\ep$ dimensions, we obtain
\begin{equation}
L_{\rm phi}(\ep,\de)=-\frac{6(1-\ep)}{(1-\ep-\de)(2-\ep-\de)(3-\ep-\de)}
L_{\rm QFT}(\ep,\de)\label{PHI}
\end{equation}
after removing irrelevant multiples of the couplings.

Thus one chooses a theory, or toy, by setting the {\tt model} switch to
one of the 10 values $\{{\tt bphz,hq,qft,mzv,hqet,hqf,hql,qcd,yuk,phi}\}$,
corresponding to the processes~(\ref{BPHZ}--\ref{PHI}).

\subsection{A variety of schemes}

We shall investigate the 10 models in 9 renormalization schemes.
This multiplicity of schemes results from a pair of ternary switches,
{\tt mscheme} and {\tt gscheme}, each of which may take the values
0, 1, or 2.

\subsubsection{Momentum, minimal and nonminimal schemes}

Our first ternary switch, {\tt mscheme}, chooses between the MOM scheme,
the MS scheme, and a nonminimal scheme (NMS).

With {\tt mscheme:=0}, we retain the full $\ep$ dependence of the bare
diagrams and generate counterterms by replacing the external momentum by a
fiducial momentum.

With {\tt mscheme:=1}, we retain in the counterterms
only the singular terms of the Laurent expansion,
after nullifying dimensionful parameters.

With {\tt mscheme:=2}, we retain
the finite term as $\ep\to0$, in addition to the singular MS terms.
Such finite counterterms occur in the delicate handling of
$\gamma_5$ in dimensional regularization~\cite{BG,BKat}.
Like the MOM and MS schemes, this NMS scheme satisfies~(\ref{rprod}).

\subsubsection{G schemes}

In~\cite{GPXT} it was observed that the appearance of $\pi^2$ (but not
of $\pi^4$) may be suppressed, in covariant field theory, by
absorbing suitable $\Gamma$ functions into the $D$-dimensional coupling,
which already absorbs the universal factor $1/(4\pi)^{D/2}$.
Our second ternary switch, {\tt gscheme}, reflects this freedom.

With {\tt gscheme:=0}, we leave the master function (\ref{BPHZ})
as it stands and modify~(\ref{HQ},\ref{QFT}) only by
including the canonical $\overline{\rm MS}$ factor $\exp(\gamma\ep)$,
which suppresses Euler's constant, $\gamma$.

With {\tt gscheme:=1}, we divide the master functions~(\ref{BPHZ}--\ref{QFT})
by their values at $\de=\ep$ and unit scale,
thus obtaining $B_1(\ep)=1$ for the one-loop
diagram, when the scale --- i.e.\ $\lambda$ in~(\ref{BPHZ});
$-2\omega$ in~(\ref{HQ}); or $p^2$ in~(\ref{QFT}) --- is set to unity.
This suppresses $\gamma$ in the HQ and QFT cases, and also
$\zeta(2)=\pi^2/6$ in the QFT case. We know from~\cite{EFT}
that $\pi^2$ is intrinsic to HQET counterterms.

With {\tt gscheme:=2}, we leave the master function (\ref{BPHZ})
as it stands and multiply~(\ref{HQ},\ref{QFT}) by $\Gamma(1-\ep)$.
This likewise suppresses $\gamma$ in the HQ and QFT cases, and also
$\zeta(2)=\pi^2/6$ in the QFT case, yet it leaves a nontrivial
value for the one-loop term $B_1(\ep)$ and hence seems
less contrived than the G scheme in~\cite{GPXT}.

We believe that the 9 schemes entailed by our pair of ternary switches
subsume much of current practice in pQFT. We make no attempt to
implement dimensional reduction, analytic regularization, or
differential renormalization, though these appear to present no
problem of principle to the global description of~\cite{DK,CK}.

\subsection{Computational strategy}

There are $\sum_{n=1}^{12}N(n)=7,813$
undecorated diagrams with $n\le12$ loops.
To obtain their renormalized values we must compute
$\sum_{n=1}^{12}2^n N(n)=24,213,878$ BPHZ terms.
To do this in 10 models and 9 renormalization schemes would require
us to process more than $2\times10^9$ BPHZ terms, with $n$th order
polynomials of a logarithm appearing at $n$ loops.
These polynomials entail $90\sum_{n=1}^{12}2^n(n+1)N(n)=27,804,356,640$
coefficients. These coefficients contain, in general, products
of odd zeta values and powers of $\pi^2$.
The number of rational coefficients of such products is
$90\sum_{n=1}^{12}2^n N(n)\sum_{k=0}^n C(k)=335,708,683,560$,
where the multiplicity, $C(k)$, of zeta products with levels up to $k$
forms the sequence
\begin{verbatim}
            1,1,2,3,4,6,8,11,14,19,24,31,39,49,61,76
\end{verbatim}
for $0\le k\le15$, neglecting the proliferation of the irreducible
MZVs of model~(\ref{MZV}), which increases these integers to give
partial sums of the Padovan sequence~\cite{BK15}.
Many of these $3.3\times10^{11}$ rationals will
be ratios of integers with $O(10)$ decimal digits. From this,
it is apparent that a complete analysis, up to 12 loops,
might entail processing several terabytes\footnote{For loop
numbers $n\le15$, we would be talking of petabytes.}
of exact integer data.
Thus computational efficiency is at a premium.

Clearly, the best strategy for any computation that involves all
7,813 diagrams is an Aufbau, in which results at $n$ loops
are held in core memory, and then used at $n+1$ loops, with only a single
new recursion of the algorithms~(\ref{D},\ref{S},\ref{brec}) for
the coproduct, antipode and bare diagrams. To achieve a
12-loop renormalized result, one needs to hold 13 terms from each
Laurent expansion, at every loop number $n<12$,
but need not hold the results for the majority of diagrams,
namely those 4,766 with $n=12$, if only weighted totals, such as
those presented in sect.~5, are required as output.

\subsection{Results and timings}

As benchmarks, we present timings for obtaining the renormalized values
for all diagrams up to a given loop number.
As interesting weights, for the output of results,
we choose those suggested by the work
of Connes and Moscovici~\cite{CM} (CM) and derived in sect.~5.
We emphasize that these weighted results were obtained
by summing over the results in the full Hopf algebra of undecorated
trees, not by specializing to the CM case {\em ab initio}.

Allocating 128MB of core memory to {\sc Reduce} on a 533MHz DecAlpha
machine we computed the case {\tt model:=bphz; mscheme:=0; gscheme:=0}.
The 10-loop CM-weighted counterterm, at scale $\lambda=1$, has finite part
\begin{equation}
S_{10}^{\rm CM,finite}=-\frac{214046911\pi^{10}}{2112}\label{sf10}
\end{equation}
obtained in less than 2 minutes. The 12-loop result
\begin{equation}
S_{12}^{\rm CM,finite}=-\frac{43556707893701\pi^{12}}{1048320}\label{sf12}
\end{equation}
took less than 2 hours. We would be interested to learn of any method that
can obtain this result, from explicit computation of all 7,813 counterterms
at loop numbers $n\le12$, in a significantly shorter time.

The next step was to test the code against the MOM-scheme OPE~(\ref{ope}),
which was verified, as were the MS-scheme OPE~(\ref{MSope}) and its
NMS-scheme version
\begin{equation}
\Gamma_{\rm NMS}(a,L_\mu-L_1)=
\mbox{$\sum_k$}
\Gamma_{\rm NMS}(a_k^{(1)},L_\mu)
\Gamma_{\rm MOM}(a_k^{(2)},-L_1)\label{NMSope}
\end{equation}
Thus we may economize, by computing only MOM-scheme results at
arbitrary momentum, while $p^2=\mu^2$ suffices in the MS and NMS cases,
thanks to the renormalization group. This helps to keep the memory
requirements within practical bounds, since the MOM-scheme renormalized
Green-function contributions are clearly rational polynomials of a log,
and are independent of the {\tt gscheme} switch.

The modifications of spin and gauge-dependence,
in~(\ref{HQET}--\ref{PHI}), introduce no new analytic feature; they merely
proliferate terms by mixing products of zeta values with different levels.
Hence we were content to compute theories~(\ref{HQF}--\ref{PHI}) to 10 loops.
We omitted the arbitrary-gauge case~(\ref{HQET}), which generates
polynomials of degree $n$ in $\chi$ at $n$ loops.
We postpone presentation of results
for the MZV model~(\ref{MZV}) to a later publication, which will address
the associated knot theory~\cite{BKP,BDK,BGK,BK15}.

At 11 loops, we encountered no difficulty in processing
models~(\ref{BPHZ}--\ref{QFT}); at 12 loops, 128MB of core memory was
sufficient only to process model~(\ref{BPHZ}).
In Table~1 we present the times (to the nearest minute)
taken to compute all the counterterms and renormalized values
up to the specified loop number, at $p^2=\mu^2$ in the MS scheme,
with {\tt gscheme=1}, in 8 models.
All timings refer to
{\sc Reduce3.5}, which was allocated 128MB of core memory on a
533MHz DecAlpha; changes of scheme have little effect on them.
Results for the CM-weighted counterterms and
renormalized Green functions are in the 12 files {\tt ha<model><loops>.out}.
In {\tt haphi10.out} one encounters, {\em inter alia},
a 25-digit integer with a 22-digit prime factor.

{\bf Table~1:} Time taken to compute all MS
counterterms and renormalized values.
$$\begin{array}{|l|cccccccc|ccc|c|}\hline{\tt model}&
{\tt bphz}&{\tt hq}&{\tt qft}&
{\tt hqf}&{\tt hql}&{\tt qcd}&{\tt yuk}&{\tt phi}&
{\tt bphz}&{\tt hq}&{\tt qft}&
{\tt bphz}\\\hline
\mbox{loops}&10&10&10&10&10&10&10&10&11&11&11&12\\
\mbox{minutes}&3&7&5&15&15&11&10&14&17&39&31&124\\\hline\end{array}$$

\section{Connes-Moscovici weights}

In~\cite{CK} an important connection was found between the
Hopf algebra ${\cal H}_T$ of the diffeomorphism group, studied
by Connes and Moscovici~\cite{CM},
and the Hopf algebra ${\cal H}_R$ of undecorated rooted trees~\cite{DK},
for which we now have 10 representations, generated
by~(\ref{BPHZ}--\ref{PHI}). The commutative part of ${\cal H}_T$
is a subalgebra of ${\cal H}_R$. Along with the obvious rainbow subalgebra
of~\cite{DKknot}, it exhausts~\cite{CK} the proper Hopf subalgebras
of ${\cal H}_R$. In consequence, there exist nonzero CM weights for our
7,813 diagrams, such that summing $n$-loop diagrams with these weights,
we arrive at representations of the grading of the CM subalgebra,
with results for the counterterms that we expect to be as
distinctive as for the rainbow diagrams in~\cite{DKknot}.

\subsection{Computation of CM weights}

The CM weights are defined as follows. Suppose that
one has the CM combination at loop number $n$. To generate the CM
combination at loop number $n+1$, one grows each previous tree
to form $n$ new trees, by attaching a new vertex to each of the original
vertices, in turn. The starting point, at
$n=1$, is the unique branchless rooted tree, with CM weight $W(\{1\})=1$.
It is not difficult to encode this algorithm, to obtain the
CM weights for the 7,813 trees with $n\le12$,
though our first encoding ran for more than an hour, until it had
finally distributed $11!=39,916,800$ terms
between the 4,766 CM weights of 12-loop diagrams. The results
are in the file {\tt hagenhaw.out}, with a coding of diagrams given
in {\tt hagendum.out}.

The defining algorithm for these CM weights appears to be very different
from those for the coproduct and antipode of ${\cal H}_R$.
For the latter, we could write simple algorithms
that were blind, at any given recursion, to all structure save that
immediately below the root (or head) of the tree.
By contrast, CM weights appear to feel all the way down to the last vertices
of the trees, which we refer to as feet\footnote{Recall that
a mathematical tree has, like a family tree, its root at the top.}.
An efficient head-first algorithm for the CM weight, $W(t)$,
was not easy to find. We finally achieved it, as follows.

The CM definition translates to the feet-first recursion
\begin{equation}
W(t)\Pi(t)=\mbox{$\sum_j$}W(f_j)\Pi(f_j)\label{CM}
\end{equation}
where the $f_j$ are all the trees obtained by removing a single foot
from $t$, and $\Pi(t)$ is a permutation factor, equal to the number of trees
that are indistinguishable from $t$ by permutations of branches originating
from any vertex. For example, the 12-loop Feynman diagram of Fig.~3
has a permutation\footnote{This permutation factor is
quite distinct from any field theoretic symmetry factor.} factor
$\Pi(t_{12})=2!\times3!\times3!=72$,
as is apparent from its divergence structure
\begin{equation}
t_{12}:=\{1,\{1,\{1\},\{1\},\{1\}\},
\{1,\{1\},\{1\},\{1\}\},\{1,\{1,\{1\}\}\}\}\label{t12}
\end{equation}
illustrated in Fig.~4.

\bookfig{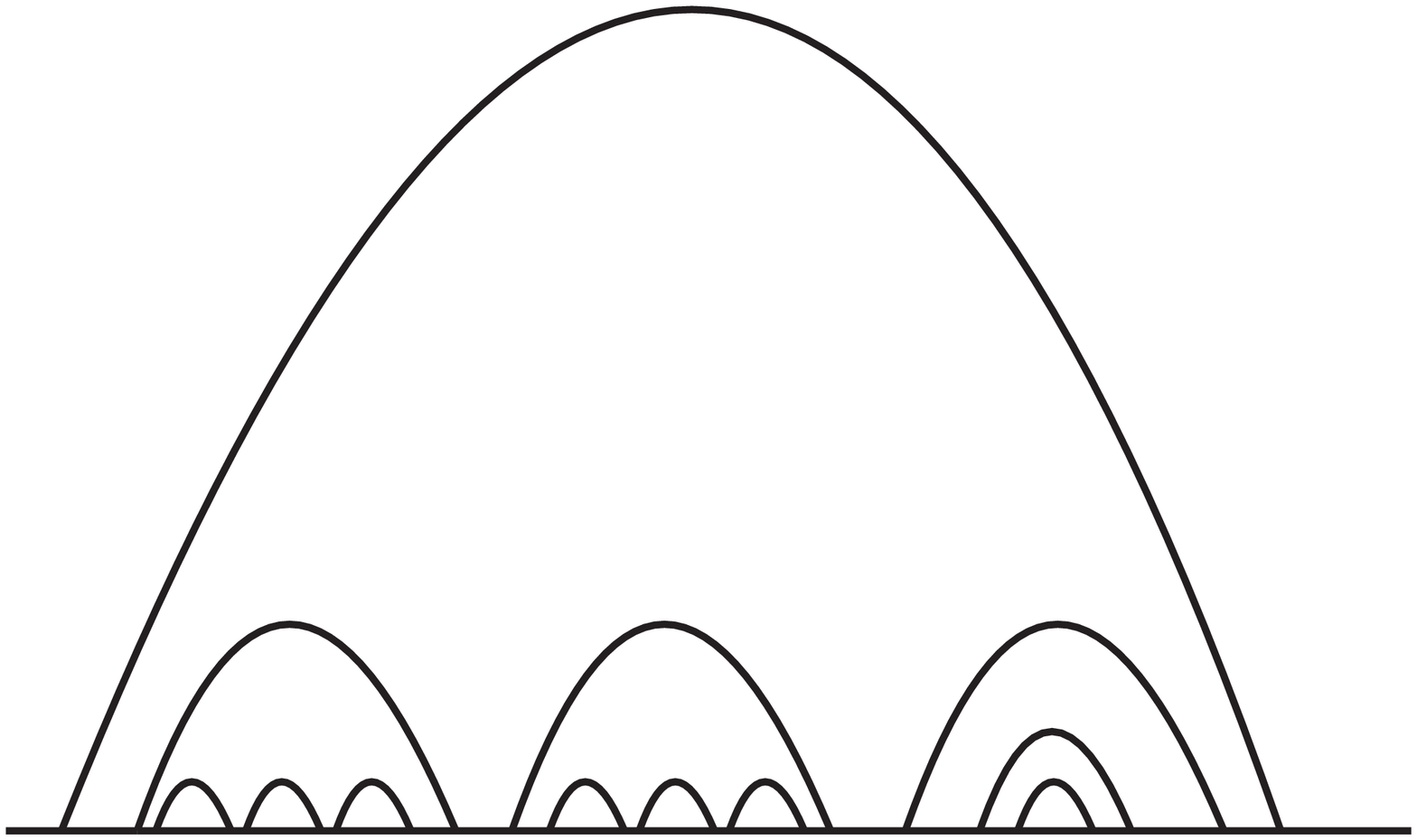}{12loops}{ddf3}{A 12-loop diagram based
on a one-loop skeleton.}{5}

\bookfig{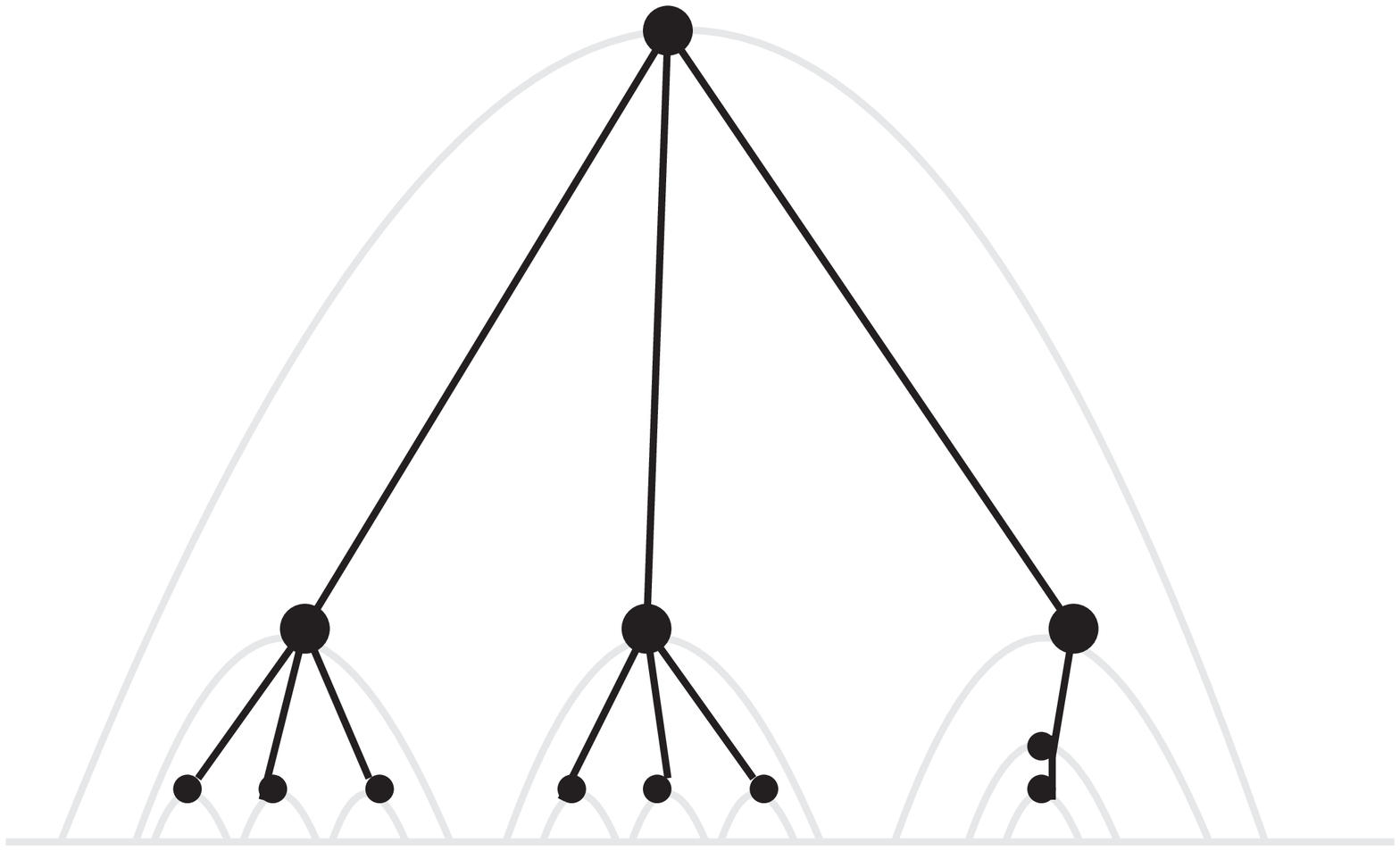}{6loops}{ddf4}{The divergence structure
of Fig.~3, exhibiting permutation symmetry.}{5}

Now the remarkable thing about the defining formula~(\ref{CM})
is that it produces no prime factor greater than $n-1$,
for a tree with $n$ vertices. This is not
at all clear from the formula, since the sum
seems to allow the possibility of adding, let us say, the integers
9 and 4, to obtain the forbidden prime 13. To show that this cannot happen,
we define another construct: the tree
factorial, $t^!$. Its recursive definition is
\begin{equation}
t^!=n\mbox{$\prod_k$}b_k^!\label{tf}
\end{equation}
where $n$ is the number of vertices of $t$ and $b_k$ are its branches.
In the pure rainbow case, where no vertex has fertility greater than unity,
$t^!$ is the ordinary factorial of the number of vertices, $n$.
In the case of Fig.~4, we have $t_{12}^!=12\times4\times4\times3!=1152$.

So now we have a feet-first sum for the CM weights~(\ref{CM}),
and a head-first product for the tree factorial~(\ref{tf}).
The trick is to convert the latter to a feet-first sum, and then express
the former as a simple head-first product.
In~\cite{OPE} it is proven, by induction, that
the tree factorials $b_k^!$ of the beheadings of any tree are related to the
tree factorials $f_j^!$ of its befootings by a wonderfully simple formula:
\begin{equation}\prod_k\frac{1}{b_k^!}=
\sum_j\frac{1}{f_j^!}\label{tff}
\end{equation}
which produces, for example, the equality
\begin{equation}
\frac{1}{4}\times\frac{1}{4}\times\frac{1}{3!}=
\frac{6}{11\times4\times3\times3!}
+\frac{1}{11\times4\times4\times2}\label{ex12}
\end{equation}
in the case of Fig.~4. Then a second induction gives
\begin{equation}
W(t)=\frac{n!}{t^!\Pi(t)}\label{CMis}
\end{equation}
with $t^!$ given by~(\ref{tf}),
and $\Pi(t)$ likewise given by a head-first recursion, namely
\begin{equation}
\Pi(t)=\pi(t)\prod_k\Pi(b_k)\label{pf}
\end{equation}
where $\pi(t)$ is the number of indistinguishable permutations
of the branches $b_k$ of $t$.

This reduces the computation of the CM weight $W(t)$
to a rather simple procedure. One associates a pair of integers to
each vertex of a tree: the first is the number of vertices of the subtree with
this root; the second is the number of indistinguishable permutations of its
branches. For a tree with $n$ vertices, one obtains $W(t)$
from~(\ref{CMis}) by dividing $n!$ by all these integers.
In the case of Fig.~4,
\begin{equation}
W(t_{12})=\frac{12!}{(12\times4\times4\times3!)\times(2!\times3!\times3!)}
=3\times5^2\times7\times11=5775\label{5775}
\end{equation}
gives the rather nontrivial Connes-Moscovici weight of the Feynman diagram
of Fig.~3. Computing the CM weights of all 4,766 12-loop
diagrams, and combining results from sect.~4 with these weights,
we expect to find something as interesting as the rational rainbow results
of~\cite{DKknot}.

\subsection{Results with CM weights}

We do indeed find rationality.
The Laurent expansion of the 12-loop CM antipode of the
BPHZ model~(\ref{BPHZ}), at $\lambda=1$, is
\begin{equation}
S_{12}^{\rm CM}=\frac{155925}{8\ep^{12}}
-\frac{43556707893701\pi^{12}}{1048320}+O(\ep^2)\label{S12}
\end{equation}
with {\em no\/} singular term that entails $\pi$. As far as Connes and
Moscovici are concerned, the Bernoulli numbers generate no
subleading Laurent residues! Moreover this depends upon no property
of the Bernoulli numbers. With arbitrary coefficients, $C_k$,
in the expansion
\begin{equation}
B_n(\ep)=\frac{1}{n\ep}\left(1+\sum_{k>0}C_k (n\pi\ep)^{2k}\right)\label{Ck}
\end{equation}
the $12$-loop CM antipode would have the same $\pi$-free singularity.
Yet the {\em finite\/} part of
\begin{eqnarray}
S_{12}^{\rm CM}&=&155925\biggl(\frac{1}{8\ep^{12}}
-\biggl\{51321600C_6+21837600C_5C_1+7001856C_4C_2+3232224C_4C_1^2
\nonumber\\&&\qquad\qquad\qquad{}
+2534490C_3^2+2208576C_3C_2C_1+229020C_3C_1^3+158336C_2^3
\nonumber\\&&\qquad\qquad\qquad{}
+146550C_2^2C_1^2+9180C_2C_1^4+\frac{231}{4}C_1^6
\biggr\}\pi^{12}\biggr)+O(\ep^2)\label{arb}
\end{eqnarray}
shows the nontrivial processing of coefficients
that results from the recursions for the coproduct~(\ref{D}),
the antipode~(\ref{S}), and the bare diagrams~(\ref{brec}),
in the presence of the CM weights~(\ref{CMis}).

With CM weights, the sole singular $n$-loop counterterm of the BPHZ
model~(\ref{BPHZ}) is of a rational, diagonal, quadratic form,
$W_2(n)/(-\ep)^n$, where
\begin{equation}
W_2(n):=\sum_{k=1}^{N(n)}\frac{W(t_k)}{t_k^!}
=\frac{1}{n!}\sum_{k=1}^{N(n)}W(t_k)\Pi(t_k)W(t_k)
\stackrel{?}{=}\frac{(n-1)!}{2^{n-1}}
\label{W2}
\end{equation}
posits a closed form, for the sum over all trees $t_k$ with $n$ vertices,
which we have verified for $n\le12$, yet have not proved in general.
Moreover, we found that $W_2(n)/(-\ep)^n$ is also the sole
$n$-loop MS counterterm in the CM-weighted QFT model
of~(\ref{QFT}) at $n\le11$ loops.
The minimal subtraction scheme is such common
practice in pQFT because it retains in counterterms only those
(products of) zeta values that are strictly necessary to
render the renormalized
Green functions finite. CM weights combine
Feynman diagrams so as to annihilate all zeta values
in model~(\ref{QFT}), just as they
annihilate powers of $\pi^2$ in model~(\ref{BPHZ}).

Inspection of~(\ref{BPHZ}) and~(\ref{QFT})
shows that each is an even function of $\de$. Thus we investigated
the most general Ansatz with this property, namely
\begin{equation}
B_n(\ep)=\frac{1}{n\ep}\sum_{j,k\ge0}C_{j,k}(n\ep)^{2j}\ep^k
\label{Ansatz}
\end{equation}
The MS counterterm from an individual 10-loop diagram involves
up to 302 terms, formed from products of $\{C_{j,k}\mid0\le2j+k<10\}$.
It took an hour to compute all 719 10-loop counterterms, with symbolic
values for the coefficients in~(\ref{Ansatz}).
Combining them with CM weights,
we annihilated 301 products of the arbitrary coefficients,
and were left with only the expected multiple,
$W_2(10)=9!/2^9=2835/4$, of $(-C_{0,0}/\ep)^{10}$.

We find it significant that the ineluctable divergences
of renormalization are so responsive to ideas from noncommutative geometry.
This strengthens our opinion that renormalization
is of real interest, in its own right, and far from
being a cause for regret.

\section{Conclusions and prospects}

In this paper, we automated the process of renormalization.
Its Hopf algebra structure, imposed by the necessity to generate
local counterterms, was summarized in a few lines of code.
The generality of this algebraic structure extends to the renormalization
of realistic particle physics problems, as in the Standard Model.
Here, we computed deep into the perturbation expansion in cases where
the remaining analysis was straightforward.

Different renormalization schemes and changes of scale were
implemented with ease, thanks to the convolutions~(\ref{ope},\ref{MSope}).
We regard no renormalization scheme as better defined than others;
all can be treated on the same algebraic footing.
Indeed, it will be shown elsewhere~\cite{OPE,DelKr}
that a minimal subtraction scheme can, for example,
be obtained as a BPHZ scheme, whose renormalization scales
are indexed by rooted trees.

The results in sect.~5 indicate a deep connection between QFT and
noncommutative geometry.
There is an index-theoretic flavour, in the annihilation of zeta values
by Connes-Moscovici weights, which deserves further study.
In particular, the new relation~(\ref{CMis}), between
tree factorials and CM weights, will be of great help in making the
connection between counterterms and diffeomorphisms, as was envisaged
in~\cite{CK} and will be made more precise in~\cite{CKnew}.
We also hope that the asymptotic enumeration of sect.~4.1 will help
Lie-algebra specialists to solve the outstanding problem of identifying
the dual of ${\cal H}_R$.

Our construction of a renormalization engine demonstrates that
the remaining challenges in computational pQFT lie in two related areas:
elucidation of algebraic relations~\cite{4TR,4TT} between decorations;
analysis of those iterated integrals~\cite{OPE},
or multiple sums~\cite{sixth},
that are the concrete representations of the truly primitive elements
which survive this filtration.

We envisage that both endeavours will be illuminated by an
interplay~\cite{CK} with noncommutative geometry~\cite{CM}.

\noindent{\bf Acknowledgments:} We thank the ESI for generous hospitality
in Vienna; Alain Connes, Victor Kac and Ivan Todorov for discussions there;
Bob Delbourgo and John Gracey for collaborations that contributed to this
work; and HUCAM for grant CHRX--CT94--0579, which contributed materially
to our meeting over the last 4 years.

\newpage\raggedright

\end{document}